\documentclass[twocolumn,prl,aps,showpacs,amsmath,amssymb]{revtex4}

\usepackage{graphicx}
\usepackage{dcolumn}
\usepackage{bm}

\begin{document}


\title{ Solution of the Two-Channel Anderson Impurity Model \\
        -- Implications for the Heavy Fermion UBe$_{13}$ -- }

\author{C. J. Bolech} 
\author{N. Andrei} 
\affiliation{Center for Materials Theory, Serin Physics Laboratory, Rutgers University\\ 
             136 Frelinghuysen Road, Piscataway, New Jersey 08854-8019, USA}

\date{December 21, 2001}

\begin{abstract} 
We solve the two-channel Anderson impurity model using the
Bethe-Ansatz. We determine the ground state and derive the
thermodynamics, obtaining the impurity entropy and specific heat over
the full range of temperature. We show that the low temperature
physics is given by a line of fixed points decribing a two-channel non
Fermi liquid behavior in the integral valence regime associated with
moment formation as well as in the mixed valence regime where no
moment forms.  We discuss relevance for the theory of UBe$_{13}$.
\end{abstract}

\pacs{75.20.Hr, 71.27.+a, 72.15.Qm}


\maketitle



In recent years a large number of alloys have been observed to deviate
from the Fermi liquid behavior, with the low temperature
thermodynamics and transport physics described by logarithmic or
fractional power laws. Among these are heavy fermion materials based
on Ce$^{3+}$ or U$^{4+}$ ions containing inner shell $f$-electrons
that do not delocalize\cite{cox98}. We shall concentrate henceforth on
U based materials, in particular UBe$_{13}$. Hund's rules and
spin-orbit coupling in the presence of a cubic crystalline electric
field lead to modeling of the U ion by a $\Gamma _{6}$ Kramers doublet
in a $5f^{3}$ configuration to be represented by the fermionic
creation operator $f_{\sigma }^{\dagger}$ and a quadrupolar
(non-magnetic) doublet $\Gamma _{3}$ in $5f^{2}$ configuration
represented by the operator $b_{\bar{\alpha }}^{\dagger }$. The
doublets hybridize with conduction electrons in $\Gamma _{8}$
representation carrying both spin ($\sigma =\uparrow ,\downarrow $)
and quadrupolar ($\alpha =\pm 1$) quantum numbers.  Strong Coulomb
repulsion requires single occupancy of the localized levels,
$f_{\sigma }^{\dagger }f_{\sigma }+b_{\bar{\alpha}}^{\dagger
}b_{\bar{\alpha}}=1$. The resulting hamiltonian is the two channel
Anderson impurity
model\cite{cox98,cox87,ramirez94,schiller98,kroha98},
\begin{eqnarray*}
H &=&H_{\text{bulk}}+H_{\text{imp}}+H_{\text{hybr}} \\
H_{\text{bulk}} &=&\int \psi _{\alpha \sigma }^{\dagger }\left( x\right)
\,\left( -i\partial _{x}\right) \,\psi _{\alpha \sigma }\left( x\right) \,dx
\\
H_{\text{imp}} &=&\varepsilon _{s}\;f_{\sigma }^{\dagger }f_{\sigma}+
\varepsilon _{q}\;b_{\bar{\alpha}}^{\dagger }b_{\bar{\alpha}} \\
H_{\text{hybr}} &=&V\;\left[ \psi ^{\dagger}_{\alpha \sigma }\left( 0\right)
b_{\bar{\alpha}}^{\dagger }f_{\sigma }+\psi _{\alpha \sigma }\left( 0\right)
f_{\sigma }^{\dagger }b_{\bar{\alpha}}\right] 
\end{eqnarray*}
The spectrum is linearized around the Fermi level and the Fermi
velocity is set to one with the resulting density of states being
$\rho=1/(2\pi )$. We shall study the model in the grand canonical
ensemble with the chemical potential $\mu$ coupled to the total number
of electrons, $N=\int \psi_{\alpha \sigma }^{\dagger }\psi_{\alpha
\sigma }\,dx + f_{\sigma }^{\dagger} f_{\sigma }$. The magnetic and
quadrupolar impurity doublets are at energies $\varepsilon_{s}$ and
$\varepsilon_{q}$ respectively. The last term gives the hybridization
of the host with the impurity. The bar over $\alpha $ indicates that
the index transforms according to the conjugate representation.

The model has been intensely studied recently via a new Monte Carlo
method \cite{schiller98} and by conserving slave boson theory
\cite{kroha98}. However, many open questions remain. In this letter we
shall present the solution of the model via a Bethe-Ansatz
construction, applicable in the past only to the single channel case
\cite{tsvwig}. We shall give a complete determination of the energy
spectrum and the thermodynamics, allowing us to follow the evolution
of the impurity from its high temperature behavior with all four
impurity states being equally populated down to the low energy
dynamics characterized by a line of fixed point hamiltonians
$H^*(\epsilon, \Delta)$ where $\varepsilon =\varepsilon
_{s}-\varepsilon _{q}$ and $\Delta=\pi\rho V^2$ (we shall hold
$\Delta$ fixed in what follows).

We shall find that the line of fixed points is characterized by a
zero-temperature entropy $S^0_{\text{imp}}=k_B\ln \sqrt{2}$ and a
specific heat $C_{\text{v}}^{\text{imp}}\sim T\ln T$ typical of
the 2-channel Kondo fixed point. However the physics along the line
varies with $\varepsilon$. Consider $n_{c}=\left\langle f_{\sigma
}^{\dagger }f_{\sigma }\right\rangle $, the amount of charge localized
at the impurity. For $\varepsilon\alt\mu-\Delta$, we find $n_c\approx
1$ signaling the magnetic integral valence regime. At intermediate
temperatures a magnetic moment forms which undergoes frustrated
screening as the temperature is lowered, leading to zero-temperature
anomalous entropy and anomalous specific heat. For
$\varepsilon\agt\mu+\Delta$ it is the quadrupolar integral regime and
a quadrupolar moment forms. In the mixed valence regime,
$|\varepsilon-\mu|\alt\Delta$, similar low temperature behavior is
observed though without the intermediate regime of moment
formation. In more detail, each point on the line of fixed points is
characterized by two energy scales, $T_{l,h}(\varepsilon)$. These
scales describe the quenching of the entropy as the temperature is
lowered: the first stage taking place at the high temperature scale
$T_{h}$, quenching the entropy from $k_B\ln 4$ to $k_B\ln 2$, the
second stage at $T_{l}$, quenching it from $k_B\ln 2$ to $k_B\ln
\sqrt{2}$. In the integral valence regime,
$|\varepsilon-\mu|\gg\Delta$, the two scales are well separated and as
long as the temperature falls between these values a moment is
present, (magnetic or quadrupolar depending on the sign of
$\varepsilon-\mu$), manifested by a finite temperature plateau
$S_{\text{imp}}=k_B\ln2$ in the entropy. It is quenched when the
temperature is lowered below $T_{l}$, with $T_{l} \to T_{K}$ in this
regime. For $\varepsilon=\mu$ the scales are equal,
$T_{l}(\mu)=T_{h}(\mu)$, and the quenching occurs in a single stage.
In a subsequent paper\cite{johannesson} we shall present the line of
boundary conformal field theories corresponding to our Bethe-Ansatz
solution, and show that the two scales parameterize the approach to
the fixed points.

The Bethe-Ansatz wave functions consist of plane waves with momenta
$\{k_{j}\}$, and amplitudes that are connected by a set of scattering
matrices (S-matrices) derived from the Hamiltonian and obeying
Yang-Baxter consistency conditions. The electron-impurity scattering
matrix is,
\begin{eqnarray*}
{\bf S}_{1,0}={\bf I}-\frac{i2\Delta }{\left( k_{1}-\varepsilon
\right) +i2\Delta }{\bf Q}_{1;0}
\end{eqnarray*}
where the index zero denotes the impurity. The operator ${\bf
Q}$ acts on quadrupolar (or flavor) space as an
``annihilation-creation'' operator, $\left[ {\bf Q} \right] _{\alpha
;\beta }^{\alpha ^{\prime };\beta ^{\prime }}=\delta _{\alpha ;\beta
}\delta ^{\alpha ^{\prime };\beta ^{\prime }}$. This S-matrix is
unitary and group invariant. To find the electron-electron S-matrix we
solve the two particle problem and find a matrix equation constraining
the possible form of the S-matrix,
\begin{eqnarray*}
 {\left({\bf S}_{2,0}-{\bf I} \right)}{\left({\bf S}_{1,0}{\bf S}_{1,2}-{\bf I} \right)} =
 {\bf P}_{1,2}^{q}{\bf P}_{1,2}^{s}
 {\left({\bf S}_{1,0}-{\bf I} \right)}{\left({\bf S}_{2,0}-{\bf S}_{1,2} \right)}
\end{eqnarray*}
where ${\bf P}^{q}$ is the flavor exchange operator and ${\bf P}^{s}$
the spin exchange operator, with $\left[ {\bf P}^{q}\right]_{\alpha
;\beta }^{\alpha ^{\prime };\beta ^{\prime}}=\delta _{\alpha }^{\beta
^{\prime }}\delta_{\beta }^{\alpha^{\prime}}$ and similarly for ${\bf
P}^{s}$. Imposing in addition the Yang-Baxter conditions, we obtain
an overdetermined set of equations admitting nevertheless a unique
solution:
\begin{eqnarray*}
{\bf S}_{1,2}=\frac{\left( k_{1}-k_{2}\right) -i2\Delta \,{\bf P}_{1,2}^{s}}
{\left( k_{1}-k_{2}\right) -i2\Delta }\frac{\left( k_{1}-k_{2}\right)
+i2\Delta \,{\bf P}_{1,2}^{q}}{\left( k_{1}-k_{2}\right) +i2\Delta }
\end{eqnarray*}

The Yang-Baxter conditions and the single occupancy constraint allow
the consistent generalization of the solution to an arbitrary number
of electrons, $N$. Imposing periodic boundary conditions, one is led
via standard methods to Bethe-Ansatz equations (BAE) that allow the
determination of the momenta and hence the spectrum: $E=\sum_{j=1}^N
k_{j}$. The BAE are
\begin{eqnarray*}
e^{ik_{j}L}= &&\prod_{\gamma =1}^{M^{s}}e_{1}\left( k_{j}-\Lambda _{\gamma
}^{s}\right) \prod_{\delta =1}^{M^{q}}[e_{1}\left( k_{j}-\Lambda _{\delta
}^{q}\right)]^{-1} \nonumber \\
-\prod_{\delta =1}^{M^{s}}&& e_{2} \left( \Lambda _{\gamma }^{s}-\Lambda
_{\delta }^{s}\right) =\prod_{j=1}^{N}e_{1}\left( \Lambda _{\gamma
}^{s}-k_{j}\right) \nonumber \\
-\prod_{\delta =1}^{M^{q}}&& e_{2} \left( \Lambda _{\gamma }^{q}-\Lambda
_{\delta }^{q}\right) =e_{1}\left( \Lambda _{\gamma }^{q}-\varepsilon
\right) \prod_{j=1}^{N}e_{1}\left( \Lambda _{\gamma }^{q}-k_{j}\right) \nonumber
\end{eqnarray*}
where $e_{n}\left( z\right) =\left( z-in\Delta \right) /\left(
z+in\Delta \right)$. $M^{s}$ is the number of down spins and the
spin-rapidities $\Lambda _{\gamma }^{s}$ describe their dynamics. The
same role is played by $M^{q}$ and the flavor rapidities $\Lambda
_{\gamma }^{q}$. The momenta $\{k_{j}\}$ (charge-rapidities)
analogously describe the charge dynamics.

Analyzing the equations in the thermodynamic limit we find that the
solutions fall into four classes: (i) real charge-rapidities; (ii)
charge-spin strings: $k_{\pm }=\Lambda^{k,s}\pm i\Delta$; (iii)
spin-strings: $\Lambda_{n}^{s\left( m_{s}\right) }= \Lambda ^{\alpha
\left( m_{s}\right) }+i\Delta\left( 1+m_{s}-2n\right)$,
$n=1,\ldots,m_{s}$ (with arbitrary length $m_{s}$); (iv)
flavor-strings: $\Lambda_{n}^{r\left( m_{q}\right) }= \Lambda ^{\alpha
\left( m_{q}\right)}+i\Delta \left( 1+m_{q}-2n\right)$,
$n=1,\ldots,m_{q}$ (with arbitrary length $m_{q}$).  A generic
full-solution consists of a combination of all four kinds.

Identifying the ground state and the excitations, we sum over the
latter to derive the free energy. It is expressed in terms of an
infinite set of functions of the variable
$\lambda=\frac{\pi}{2\Delta}(k-\mu)$. These functions, $\{
\eta_{b}(\lambda), \eta_{u}(\lambda), \eta_{s}^{n}(\lambda),
\eta_{q}^{n}(\lambda) \}_{n=1,\ldots ,\infty}$, determine the
distribution probabilities of the various excitations at temperature
$T$ (entering the equations as $\tau=\frac{\pi T}{4\Delta}$): the
charge bound states, charge unbound states, spin excitations and
flavor excitations, respectively. These distribution functions satisfy
an infinite set of coupled non-linear integral equations:
the thermodynamic Bethe Ansatz (TBA) equations,
\begin{eqnarray*}
\ln \bar{\eta}_{b} &=&-\frac{\lambda}{\tau}+
                      G*\left[ {\rm f}_{q2}-{\rm f}_{\bar{u}} \right] 
\\
\ln \bar{\eta}_{u} &=&-\frac{\lambda}{2\tau}+
                      G*\left[-{\rm f}_{\bar{b}}+{\rm f}_{s1}+{\rm f}_{q1}\right] 
\\
\ln \eta _{sn} &=&G*\left[ \delta _{n,1}{\rm f}_{\bar{u}}
                          +\bar{\delta}_{n,1}{\rm f}_{sn-1}
                          +{\rm f}_{sn+1}\right]
\\
\ln \eta _{qn} &=&G*\left[-\delta _{n,1}{\rm f}_{\bar{u}}
                          -\delta_{n,2}{\rm f}_{\bar{b}}
                          +\bar{\delta}_{n,1}{\rm f}_{qn-1}
                          +{\rm f}_{qn+1}\right]
\end{eqnarray*}
where we have introduced the notations $\bar{\eta}=1/\eta $, ${\rm
f_{\alpha}} =\ln \left( 1+\eta_{\alpha} \right) $, ${\rm
f}_{\bar{\alpha}}=\ln \left( 1+\bar{\eta}_{\alpha }\right) $ and
$\bar{\delta}_{n,1}=1-\delta _{n,1}$. The convolutions (denoted by
$*$) are defined with the kernel $G(\lambda) = \frac{1}{2\pi \cosh
\lambda}$. The magnetic field $h_{s}$ and the quadrupolar field
$h_{q}$ determine the asymptotic conditions: $\lim_{n\rightarrow
\infty }(K_{n+1}*{\rm f}_{sn}-K_{n}*{\rm f}_{sn+1}) = -2h_{s}/T$ and
$\lim_{n\rightarrow \infty }(K_{n+1}*{\rm f}_{qn}-K_{n}*{\rm
f}_{qn+1})=-2h_{q}/T$, where
$K_{n}(\lambda)=\frac{2n}{(2\lambda)^{2}+(n\pi)^{2}}$.

In terms of the distribution functions, the free energy
$F=F_{\text{bulk}}+F_{\text{imp}}$ is given by:
\begin{eqnarray*}
F_{\text{bulk}}&=&-TL\int \rho\,{\rm f}_{\bar{u}}\;d\lambda
                  -2TL\int \rho\,{\rm f}_{\bar{b}}\;d\lambda \\
F_{\text{imp}} &=&\varepsilon _{q}
           -T\,\left[ G*\left( K_{1}*{\rm f}_{\bar{u}}+K_{2}*{\rm f}_{\bar{b}}
                              +{\rm f}_{q1}\right) \right]_{\lambda=\frac{\pi}{J}}
\end{eqnarray*}
where $J=\frac{2\Delta}{\varepsilon-\mu}$. Having solved for the
distributions for given values of temperature and fields, one can
compute the free energy of the impurity for any value of
$\varepsilon$. The bulk part of the free energy is divergent and
requires the introduction of a regularization scheme.
On the other hand, the impurity contribution to the free energy is
regular and independent of the cut-off procedure.

We turn now to study the TBA equations, considering first the zero
temperature limit for which it is convenient to introduce $\xi =\tau
\ln \eta $. We then have, $\lim_{\tau\rightarrow 0^{+}} \tau{\rm
f}=\xi^{+}$ and $\lim_{\tau\rightarrow 0^{+}} \tau{\rm\bar
f}=\xi^{-}$, where $2\xi^{\pm}=\xi\pm|\xi|$, in terms of which we are
able to solve explicitly the TBA equations. We find: $\xi_{b} =
\lambda,~ \xi_{q2} = \xi_{q2}^{-} = G*\xi_{b}^{-},~ \xi_{u} =
\xi_{u}^{+} = \frac{\lambda}{2}-\xi _{q2}$ with all the other $\xi$'s
vanishing (the result of the convolution giving $\xi _{q2}$ can be
written down in terms of dilogarithms). Thus in the absence of
external fields, the ground state is built out of a sea of charge-spin
strings filled up to $\lambda=0$ (i.e. $k=\mu $) and a completely
filled sea of 2-flavor-strings. The zero-temperature impurity level
occupancy (and therefore also the charge susceptibility) can be
deduced in a closed form: $ n_{c}^{0} =\int_{-\infty }^{+\infty}
\frac{4(\lambda-\pi/J)\,
\xi_{q2}}{[(\lambda-\pi/J)^{2}+\pi^2]^2}\,d\lambda$.  The occupancy is
integral: $n_{c}^{0} \approx 1,0$ for $|\varepsilon-\mu|\gg\Delta$,
and non-integral elsewhere.

We turn next to study the low temperature physics, $\tau\ll 1$. In the
integral valence regimes, $|\varepsilon-\mu|\gg\Delta$, $|J|\gg 1$ and
the main contribution to the free energy comes from $|\lambda| \approx
\pi/|J| \gg 1$.  It is convenient to rewrite some equations in the
TBA. Using the exact relation:
$\ln\eta_b-\ln{\bar{\eta}}_{q1}=\lambda/\tau$, we can eliminate ${\rm
f}_{\bar{b}}$ from the equations for $\bar{\eta}_{u}$ and $\eta _{q2}$
and write them as: $\ln \bar{\eta}_{u} =-E_{+}^d/\tau+G*\left[{\rm
f}_{s1}+{\rm f}_{q1}\right]$ and $ \ln \eta _{q2} =
-E_{-}^d/\tau+G*\left[ {\rm f}_{q1}+{\rm f}_{q3}\right]$ with $
E_{\pm}^d/\tau=
G*\ln\left(1+e^{\pm\frac{1}{\tau}(\lambda-\tau\ln\eta_{q1})}\right)$. The
terms $E_{\pm}^d$ become driving (inhomogeneous) terms at low
temperatures as $\tau\ln\eta_{q1}$ tends to zero (with $\tau^2$
corrections).  We can further approximate them as follows:
\begin{equation*}
E_{\pm}^d/\tau \xrightarrow[\tau\ll 1]{} G*[\lambda/\tau]^{\pm}\to
\left\{
\begin{aligned}
     \xrightarrow[\mp\lambda\agt  1]{} e^{\pm\lambda}/\pi\tau \\
     \xrightarrow[\mp\lambda\alt -1]{} \lambda/2\tau
\end{aligned}
\right.
\end{equation*}
Changing variables, $\zeta=\lambda-\frac{\pi}{J}$, the driving terms
become, in the integral valence regime:
$E_{\pm}^d/\tau=e^{\pm\zeta-\ln T/T_{\pm}}$ where $T_{\pm}=\frac{4
\Delta}{\pi^2}e^{\pm\pi/J}$. The relation of $T_{\pm}$ to $T_{l,h}$
depends on the sign of $\varepsilon -\mu$ and is discussed below.

We analyze separately the magnetic and quadrupolar moment regimes.
Consider first the magnetic case where $\varepsilon\ll\mu-\Delta$, and
therefore $\lambda\ll -1$. In the limit of small $\tau$, the driving
term $E_{-}^d$ diverges faster than $E_{+}^d$ that is compensated by a
decaying exponential in the numerator; thus $\eta_{q2}$ tends to zero
exponentially fast cutting away the equations for the higher q-flavor
$\eta$'s. After the identifications $\bar{\eta}_{q1}\rightarrow \eta
_{1}^{s}$, $\bar{\eta}_{u}\rightarrow \eta _{2}^{s}$ and $\eta
_{sn}\rightarrow \eta _{n+2}^{s}$, we recover the TBA equations of the
2--channel Kondo problem\cite{andrei87}. The Kondo temperature is
$T_K=T_l=T_{+}$.  ($T_{-}$ is large, outside the low temperature
approximation range). The identification of the resulting TBA
equations in this limit with those of the 2--channel Kondo model
indicates that at low temperatures, when $E_{-}^d/\tau$ becomes very
large, the system has a localized magnetic moment. When temperature
goes below $T_K$, overscreening will take place and in particular an
entropy $S_{\text{imp}}=k_B\ln \sqrt{2}$ will arise. This will also be
found numerically (see below).

On the other hand, when $\varepsilon\gg \mu+\Delta$, the limit of
vanishing $\tau$ drives $\bar{\eta}_{u}$ to zero exponentially fast
and cuts away the spin $\eta $'s. No remapping is required, and we
recover again the TBA equations of the 2--channel Kondo model but this
time for the quadrupolar degrees of freedom.  The relative minus sign
in the driving term reverses the `direction of lowering temperatures'
with respect to the previous case and has no further consequences. The
Kondo temperature is $T_K=T_l=T_{-}$, below which the two spin
channels will overscreen a localized quadrupolar moment.

Now we study the mixed valence regime, $|\varepsilon -\mu|\alt\Delta
$, where $n_c\approx 1/2$. As $|J|\gg 1$, we need the solutions around
$\lambda\approx 0$.  Then both driving terms are approximately
$E_{\pm}^d/\tau=e^{\pm\zeta-\ln T/T_{\pm}}$ and both diverge
simultaneously in the low temperature limit. We repeat the standard
procedure of shifting variables with $\pm\ln
T/T_K$\cite{andrei87}. Depending on the choice of sign, one of the
driving terms diverges as $1/\tau^2$ and the temperature `disappears'
from the other one. We recover one of the two cases outlined above but
this time the equations are valid only at temperatures below $T_l$ and
there is no localized moment regime. However, a residual entropy
$S_{\text{imp}}=k_B\ln \sqrt{2}$ arises also in this regime.

\begin{figure}
\includegraphics[width=0.45\textwidth]{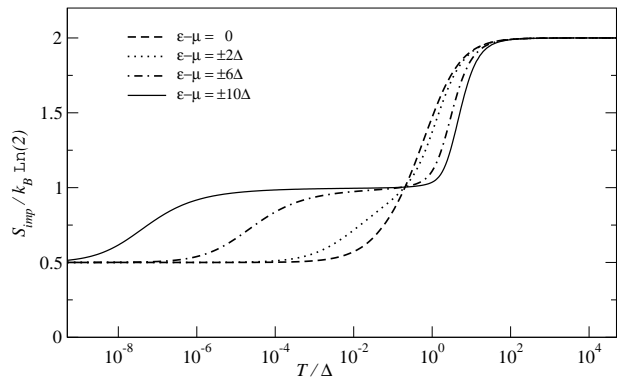}
\caption{\label{S_Plot}Impurity contribution to the entropy as a
function of temperature for different values of $\varepsilon-\mu$. As
the temperature goes to zero all curves approach the universal value
$k_B\ln\sqrt{2}$. Positive and negative values of $\varepsilon-\mu$
fall on top of each other.}
\end{figure}

Finally, we turn to the finite temperature thermodynamics. We obtain
it by solving numerically the TBA equations. In Fig.~\ref{S_Plot} we
show the behavior of the impurity entropy as a function of
temperature for different values of $\varepsilon$.  At high
temperatures the entropy is $k_B\ln4$ in agreement with the size of
the impurity Hilbert space. For $|\varepsilon-\mu| \gg \Delta$ the
impurity entropy is quenched in two stages. The degrees of freedom
corresponding to the higher  energy doublet are frozen
first. The entropy becomes $k_B\ln2$ and the system is in
a localized magnetic or quadrupolar moment regime depending on the
sign of $\varepsilon -\mu$. As the temperature is further decreased, the
remaining degrees of freedom  undergo  frustrated
screening leading to  entropy
$k_B\ln\sqrt{2}$. On the other hand, for values of $|\varepsilon -\mu|
\ll \Delta$, the quenching process takes place in a single stage. As
$\varepsilon-\mu$ is varied the behavior interpolates continuously
between the magnetic and the quadrupolar scenarios. The zero
temperature entropy is found to be independent of $\varepsilon$ in
accordance with the analytic study of the TBA equations. Note also the
presence of a crossing point, a temperature $T_{\text{cross}} \approx
0.1 \Delta$ where all entropies take the value $S_{\text{imp}}=k_B\ln
2$ independently of $\varepsilon$ (cf.~Ref.~\onlinecite{vollhardt97}).

\begin{figure}
\includegraphics[width=0.45\textwidth]{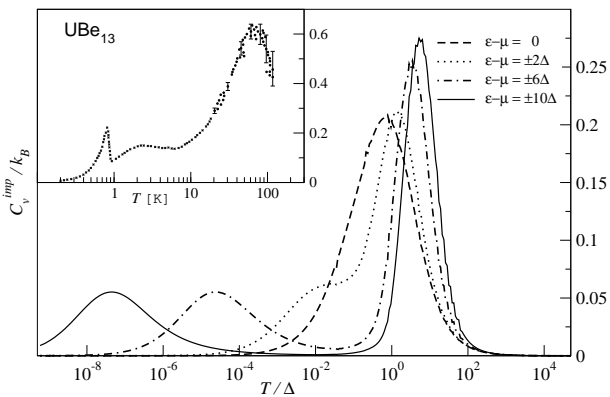}
\caption{\label{CvPlot}Main plot: impurity contribution to the
specific heat as a function of temperature for different values of
$\varepsilon-\mu$. As this parameter approaches zero the Kondo
contribution (left) and the Schottky anomaly (right) collapse in a
single peak. Inset: experimental data for the $5f$-derived specific
heat of UBe$_13$.}
\end{figure}

In Fig.~\ref{CvPlot} we show the impurity contribution to the specific
heat. The two stage quenching process gives rise to two distinct
peaks. The lower temperature peak is the Kondo contribution
centered around $T_{l}$ whereas the higher temperature peak often
referred to as the Schottky anomaly is centered around a temperature
$T_{h}$. Approximate expressions for $T_{h,l}$ can be read off from
the curves: $T_{h,l}(\varepsilon) \approx
\frac{\Delta}{a\pi^2}\ln(1+2a\,e^{\pm\frac{\pi}{2\Delta}|\varepsilon-\mu|})$
with $1<a<4$ (this expression goes over to $T_\pm$ defined before in
the appropriate limits). For large $|\varepsilon-\mu|$ the two peaks
are clearly separated and the area under the Kondo peak is
$k_B\ln\sqrt{2}$ while that under the Schottky peak  is $k_B\ln 2$.

As mentioned earlier, the model was
proposed as a description for the U ion physics of UBe$_{13}$. It is
expected to describe the lattice above some coherence temperature.  We
provide in the inset of Fig.~\ref{CvPlot} the experimental data for
the $5f$-derived specific heat of the compound. It is obtained by
subtracting from its total specific heat, the specific heat of the
isostructural compound ThBe$_{13}$ containing no $5f$
electrons\cite{felten86}. This way one is subtracting the phonon
contribution as well as the electronic contribution from electrons in
$s$, $p$ and $d$ shells (the procedure is quite involved and we refer
the reader to the articles cited just above for full details). The
sharp feature at $~0.8K$ signals the superconducting transition of
UBe$_{13}$ and falls outside the range where this compound might be
described by an impurity model.

Concentrating on the temperature range containing the Kondo and
Schottky peaks we conclude that no values of $\varepsilon$ and
$\Delta$ yield a good fit. Further, the entropy obtained by
integrating the weight under the experimental curve falls between
$k_B\ln4$ and $k_B\ln6$. This suggests that a full description of the
impurity may involve another high energy multiplet (possibly a triplet
cf.~Ref.~\onlinecite{koga99}), almost degenerate with the $\Gamma_3$
to yield a single peak for the Schottky anomaly. The nature of the
multiplet could be deduced from further specific heat
measurements. For an n-plet degenerate with the $\Gamma_3$, one has an
$SU(2)\times SU(n+2)$ Anderson model and the area under
$C_{\text{v}}^{\text{imp}}/T$ is then given by
$k_B\ln[\frac{n+4}{2}\sec\frac{\pi}{n+4}]$, while if the n-plet is
slightly split off the doublet, the area is given by
$k_B\ln[\frac{n+4}{\sqrt{2}}]$ \cite{jerez}.  In order to be certain
of having eliminated lattice effects it would be better to carry out
the measurements on U$_{1-x}$Th$_x$Be$_{13}$. For $x>0.1$ the compound
has no longer a superconducting transition and the lattice coherence
effects are largely suppressed. Further, there are several
experimental indications that support the idea of an impurity model
description of the thoriated compound for a wide range of temperatures
\cite{aliev96}.

In subsequent work we shall present the solution of the general
$SU(N)\times SU(M)$ model and study the effects of magnetic and
quadrupolar fields.  These will help identify what particular impurity
model is best fit for describing U$_{1-x}$Th$_x$Be$_{13}$
\cite{koga99}.

Part of the work was done while one of the authors (N.~A.)  was a Lady
Davis fellow at the Hebrew University in Jerusalem. He thanks the
physics department for its warm hospitality.  We are grateful to
P. Coleman, G. Kotliar, H. Kroha, H. Johannesson A. Schiller and
P. W\"olfle for illuminating discussions, and to A. Jerez,
S. Kancharla, A. Rosch and N. Shah for their comments on the
manuscript. The experimental data is reproduced from
Ref.~\onlinecite{felten86} with kind permission from the authors.

\end{document}